\documentclass[aps,prl,twocolumn]{revtex4-2}
\usepackage{mathtools}
\usepackage{xcolor}
\usepackage{graphicx}
\usepackage{amsmath}
\usepackage{amssymb}

\begin{document}

\title{Multiscale Energy Spreading in Hard-Particle Chains}


\author{Arkady Pikovsky}
 \email{pikovsky@uni-potsdam.de}
 \affiliation{Department of Physics and Astronomy, University of Potsdam, Karl-Liebknecht-Str. 24/25, 14476, Potsdam-Golm, Germany}

\date{\today}

\begin{abstract}
We consider a one-dimensional array of particles interacting via an infinite well potential. We explore the properties of energy spreading from an initial state where only a group of particles has non-zero velocities while others are resting. We characterize anomalous diffusion of the active domain via moments and entropies of the energy distribution. Only in the special cases of a single-well potential (hard-particle gas) and of the distance between the particles being half of the potential width does the diffusion have a single scale; otherwise, a multiscale anomalous diffusion is observed.
\end{abstract}

\maketitle

\textit{Introduction.}  
One-dimensional classical nonlinear lattices posses quite peculiar transport properties. On the one hand, many such systems demonstrate anomalous heat transport, in which the finite-size heat conductivity diverges as a power of the lattice length~\cite{Lepri-16,livi2023anomalous,benenti2023non}. In a typical setup, one attaches a lattice to two heat baths with different temperatures and measures the energy exchanges with them. Because according to the Green-Kubo theory, the conductivity can be expressed in terms of the equilibrium fluctuations of the heat current, anomalous transport manifests itself in a non-integrable power-law decay of the correlation function of the energy current~\cite{lepri1998anomalous}. Another manifestation of anomalous heat transport is a superdiffusive spreading of local energy perturbations on top of the equilibrium state with finite temperature (finite energy density)~\cite{cipriani2005anomalous}. Theoretical and numerical findings about anomalous heat transport have been confirmed in experiments~\cite{Chang-08,Upadhyaya-16,Crnjar2018,yang2021observation}.

Another setup where one-dimensional lattices demonstrate nontrivial transport properties is the spreading of initially localized perturbations on top of a zero-temperature state. Due to the effect of Anderson localization~\cite{anderson-10}, in a linear lattice, already a small amount of disorder leads to exponential localization of eigenmodes, which blocks spreading at large times. However, the nonlinearity of the lattice may result in chaos, which destroys localization and leads to a subdiffusive spreading. One popular example is a disordered nonlinear Schroedinger lattice~\cite{Pikovsky-Shepelyansky-08,flach2009universal,Fishman-Krivolapov-Soffer-12,flach2015nonlinear}. However, even for this widely studied model it still not clear whether spreading persist at very large times, as chaos may degenerate into localized quasiperiodic modes~\cite{Pikovsky-Fishman-11,aubry23}. Spreading in a nonlinear Schroedinger lattice has been observed in optical experiments~\cite{Lahini-09}. While in the context of the Schroedinger lattice, one follows the spreading of the wave packet, for nonlinear lattices of Klein-Gordon or Fermi-Pasta-Ulam-Tsingoi type, one studies the spreading of energy from a localized perturbation on top of a zero-temperature state~\cite{Mulansky-Ahnert-Pikovsky-11,Mulansky-Pikovsky-13,Martinez-16,PhysRevE.99.032211,Pikovsky-20}. While in many numerical experiments a subdiffusive spreading is observed, asymptotic regimes at large times remain elusive.  For the experimental realization of such a spreading in a disordered granular chain, see Ref.~\cite{kim2018direct}.

In this Letter, we explore energy spreading on top of a zero-temperature state for hard-particle models previously studied in the context of anomalous heat transport. These models share basic properties like conservation of momentum with lattices with smooth coupling potentials, but allow for an efficient numerical implementation. We will demonstrate that while in some cases anomalous superdiffusion or subdiffusion, characterized just by one exponent (mono-scale diffusion), is observed, generally a hard-particle chain demonstrates multi-scaled diffusion, where differently defined ``lengths'' of the spreading domain grow with different exponents.

\textit{Model formulation.} Our basic model is a one-dimensional chain of particles interacting via an infinite well potential~\cite{delfini2007energy,politi2011heat,mejia2019heat}. This hard-particle chain (HPC) is defined via the Hamilton function
\begin{equation}
\begin{aligned}
H&=\sum_i \frac{p_i^2}{2m_i}+U(x_{i+1}-x_i)\;,\\
U(\Delta x)&=\begin{cases} \infty & \Delta x <0 \text{ and }\Delta x >a^{-1}\;,\\ 0 & 0\leq \Delta x\leq a^{-1}\;.\end{cases}
\end{aligned}
\label{eq:1}
\end{equation}
The essential dimensionless parameter of the problem is $0\leq a<1$, the ratio of the mean distance between the particles and the potential width. We set the initial distance between the particles to $1$; thus the parameter $a$ enters the definition of the potential (the width of the well) in \eqref{eq:1}.
To ensure the dynamics is non-integrable, we set random masses according to a uniform distribution $0.5\leq m_i\leq 1.5$. 

In the case $a=0$, the double-well potential becomes a hard-core potential, and the model reduces to a famous one-dimensional hard-particle gas (HPG) model~\cite{casati1986energy,Grassberger-02,lepri2020too}, where only elastic collisions at $x_{i+1}=x_i$ happen. Remarkably, the HPC model is symmetric with respect to the change of the parameter $a\to 1-a$~\cite{mejia2019heat}. The value $a=1/2$ is special because here, the two types of the collisions at $x_{i+1}=x_i$ and at $x_{i+1}=x_i+a^{-1}$ become in equilibrium equally probably, and the pressure vanishes. Therefore, we restrict our attention below to the interval $0\leq a\leq 1/2$.

The HPC \eqref{eq:1} conserves energy and momentum, and in the heat transfer setup has demonstrated anomalous heat conductivity~\cite{delfini2007energy,politi2011heat,mejia2019heat}. Below in this letter we study energy-spreading properties at zero temperature. We start with a configuration where positions are equidistant $x_i(0)=i$, and the momenta $p_i$ of the particles are randomly set in a small domain (ten sites) around $i=0$, other particles are at rest  $p_i(0)=0$. Because of the possibility of time rescaling, without loss of generality, the total energy can be set to $1$. Furthermore, we set the total initial momentum to zero.   

\begin{figure}
\centering
\includegraphics[width=\columnwidth]{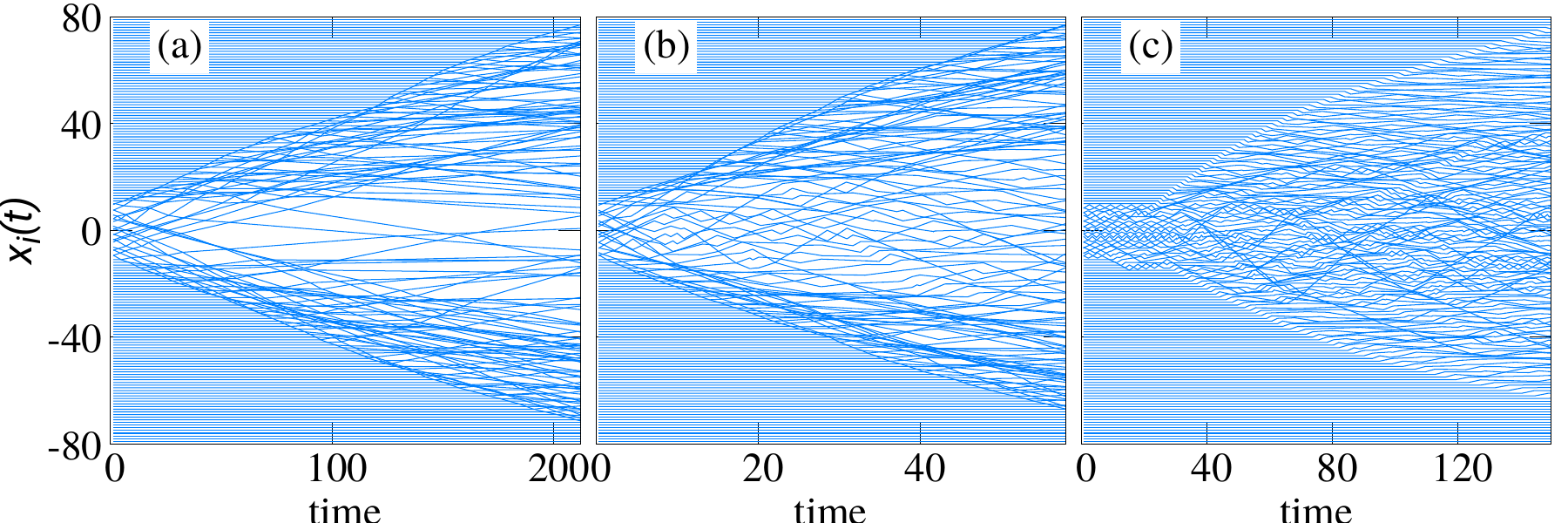}
\caption{Examples of the energy spreading process: trajectories $x_i(t)$ of particles vs time for three values of parameter $a$: panel (a) $a=0$; panel (b) $a=0.25$; panel (c) $a=0.5$. }
\label{fig:1}
\end{figure}

\textit{Energy spreading and its characterization.}
The phenomenology of the dynamics is simple (Fig.~\ref{fig:1}): more and more particles are involved in the nontrivial dynamics via collisions, and a spreading ``active domain'' consisting of particles having non-zero energy is formed.  At each time instant, only a finite number of particles belong to the active domain, and this number $L(t)$ is a natural measure of the domain width. However, this number is only one of the possible definitions of the domain width. Because we normalize the total energy to one, and the energies of the particles are non-negative, the set of local energies $E_i=p_i^2/(2m_i)$ can be interpreted as a probability distribution. Accordingly, we can characterize the distribution width using this distribution's moments or entropies.

In the former approach, we define the ``center of energy''$\langle x\rangle$, the moments $M_p$, and the corresponding moment-based sizes of the domain $\ell_p$ according to relations
\begin{equation}
\langle x\rangle=\sum_i x_i E_i,\; M_{p}=\sum_i |x_i-\langle x\rangle|^p E_i,\; \ell_p=(M_{p})^{1/p}\;.
\label{eq:mom}
\end{equation}
Here, the indices $p>0$ need not be integers. 

Another way to characterize distributions is to calculate their Renyi entropies $I_{q}$ depending on index $q\geq 0$, and to use them to define the widths $\mathcal{L}_q$:
\begin{equation}
I_{q}=\frac{\log \sum_i E_i^q}{1-q},\quad \mathcal{L}_q=\exp[I_q]\;.
\label{eq:ent}
\end{equation}
 Note that for $q=0$, the entropy is the logarithm of the support of the distribution $I_0=\log L$, thus $L=\mathcal{L}_0$. Another widely used case is $q=2$ which corresponds to the participation number, broadly utilized in the context of wave packet spreading.  

\begin{figure}
\centering
\includegraphics[width=\columnwidth]{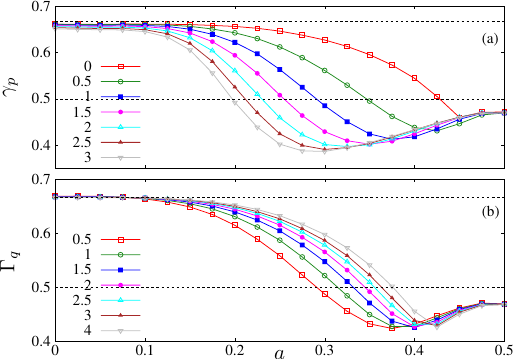}
\caption{Power law growth rates in dependence on the indices and parameter $a$. Panel (a): exponents $\gamma_p$ for indices $p$ indicated in the marker descriptions; panel (b): exponents $\Gamma_q$ for indices $q$ indicated in the marker descriptions. The dotted black lines show values $2/3$ and $1/2$. }
\label{fig:2}
\end{figure}

Plots of the lengths $\ell_p,\mathcal{L}_q$ reveal that these quantities grow as power laws: $\ell_p\sim t^{\gamma_p}$, $\mathcal{L}_q\sim t^{\Gamma_q}$ (cf. observation of index-dependent growth powers of the moments for dynamically generated diffusion processes in Refs.~\cite{aranson1990anomalous,pikovsky1991statistical}). 
We present these powers as functions of parameter $a$ in Fig.~\ref{fig:2}. 

\begin{figure}
\centering
\includegraphics[width=0.9\columnwidth]{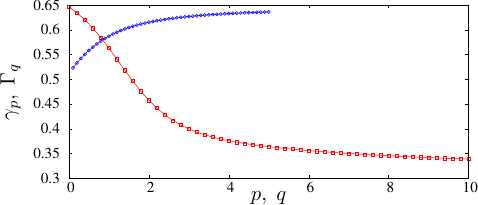}
\caption{Power laws exponents in dependence of indices for $a=0.25$. Red squares: entropies-based exponents $\Gamma_q$, blue circles: moments-based exponents $\gamma_p$.}
\label{fig:3}
\end{figure}

The main conclusion from the data of Fig.~\ref{fig:2} is that the powers for all the indices coincide if $a=0$ (pure HPG) or $a=1/2$ (HPC with zero pressure in equilibrium), so that one can speak on \textit{monoscale} spreading in these cases.  In contradistinction, in-between, there is a strong dependence of exponents $\gamma_p,\Gamma_q$ on indices $p,q$, manifesting \textit{multi-scale spreading} (in the context of diffusion processes one speaks in this case on ``strong anomalous diffusion''~\cite{castiglione1999strong}). We illustrate these dependencies in a larger range of indices for $a=1/4$ in Fig.~\ref{fig:3}. 

We stress that in all the cases, diffusive spreading of the energy is nontrivial: it is superdiffusive for small $a$ and subdiffusive for $a$ close to $1/2$.  However, for $0.2\lesssim a\lesssim 0.4$, some exponents are larger than $1/2$ and some smaller than $1/2$, as illustrated in Fig.~\ref{fig:3}. In particular, for $a=0$ (HPG case), one observes $\gamma_p\approx\Gamma_q\approx 2/3$. This power can be obtained from the following simple scaling arguments. Suppose that the energy is nearly uniformly distributed among $L$ active particles. Then, the energy of the boundary particle is $\sim L^{-1}$, and its velocity is $\sim L^{-1/2}$. The time to hit the next resting particle outside the active domain is inversely proportional to the velocity, and after this happens, the active domain increases by $1$. Thus, $dt/dL\sim L^{1/2}$ and solving this equation we get $t\sim L^{3/2}$. This yields the scaling law for the domain spreading $L\sim t^{2/3}$.

\textit{Scaling properties of the distributions.} Next, we describe the scaling properties of the distributions of the active particles. We introduce the coarse-grained density of the particles and two coarse-grained distributions of the energy: energy per particle distribution and energy density. 
To scale these distributions, we notice that $L(t)$ is the length of the active domain and the number of active particles (because in our setup, the spacing is $1$). So we introduce the normalized particle index $\nu=(i-i_{left})L^{-1}$, $0\leq \nu\leq 1$, and the normalize particle position
$\xi=(x_i-x_{i_{left}})L^{-1}-0.5$, $-0.5\leq \xi \leq 0.5$. Here, $i_{left}$ is the index of the left-most active particle (left border of the active domain). Thus, plotting $i$ vs $x_i$ we obtain a cumulative distribution of particles in scaling coordinates $\nu(\xi)$, and its derivative yields the particle density $\rho(\xi)=\frac{d\nu}{d\xi}$. Similarly, we introduce the cumulative energy as
$\epsilon_i=\sum_{k=i_{left}}^i E_k$, so that $0\leq \epsilon\leq 1$. By plotting $\epsilon_i$ vs $i$ we obtain a curve $\epsilon(\nu)$ which yields the density of energy per particle $W(\nu)=\frac{d\epsilon}{d\nu}$. The density of energy $w(\xi)=\frac{d\epsilon}{d\xi}$ can be obtained from the obvious relation $w(\xi)=W(\nu)\rho(\xi)$.

To distinguish mono-scale and multi-scale regimes, we must compare densities occurring at different sizes $L$ of the active domain. If these densities coincide, then the length $L$ delivers a complete characterization of the distributions, and all the powers $\gamma_p,\Gamma_q$ are equal. Multi-scaling occurs if the densities at different values of length $L$ have different shapes. 

\begin{figure}
\centering
\includegraphics[width=\columnwidth]{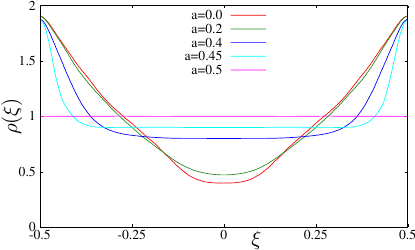}
\caption{Densities $\rho(\xi)$ for different values of parameter $a$.}
\label{fig:4}
\end{figure}

The first observation from the numerics is that the renormalized particle density $\rho(\xi)$ does not depend on length $L$, for sufficiently large $L$. The profiles at different parameter values $a$ are shown in Fig.~\ref{fig:4}. Notice that the density outside the active domain is $\rho=1$. For $a$ close to zero, two steps of size $\approx 0.8$ are formed at the borders of the active domain, and these ``shocks'' propagate superdiffusively. For larger $a$, the distribution in the bulk becomes nearly flat, and the dense shock regions become thinner. Finally, at $a=0.5$, the shocks disappear, and the density over the active domain is uniform $\rho=1$.

\begin{figure}
\centering
\includegraphics[width=\columnwidth]{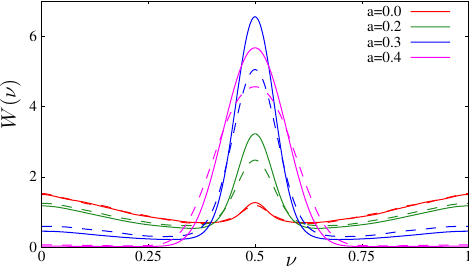}
\caption{Densities of energy per particle $W(\nu)$ for different values of parameter $a$. Full lines: at $L=4\cdot 10^4$, dashed lines: at $L=10^4$.}
\label{fig:5}
\end{figure}

Next, we discuss the energy distributions $W(\nu)$ and $w(\xi)$. In Fig.~\ref{fig:5}, we show the coarse-grained energy-per-particle distribution densities at several values of parameter $a$ at two different instants of time: one at which the total width of the active domain is $L=10^4$ and another one at which $L=4\cdot 10^4$. One can see that for $a=0$, these densities practically coincide, while for other values of $a$, they are significantly different. This corresponds well with multi-scaling at these values of $a$. The energy-per-particle distributions possess a ``hot core'' of locally highly energetic particles at the center of the active domain. This is accompanied by tails of particles with relatively low energy; in these tails, however, energy grows toward the edges of the active domain.

\begin{figure}
\centering
\includegraphics[width=\columnwidth]{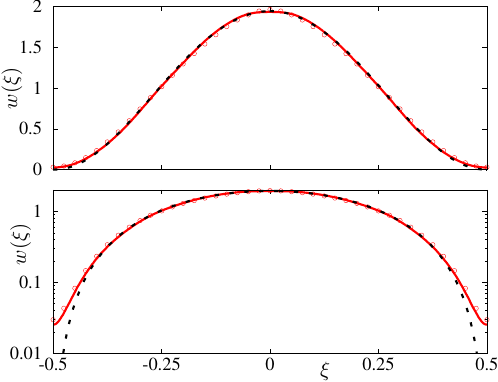}
\caption{Density of energy $w(\xi)$ for $a=1/2$, in the linear and in the logarithmic formats. Full lines: at $L=4\cdot 10^4$, red circles: at $L=2\cdot 10^4$. Black dashed line: the fit discussed in the text.}
\label{fig:6}
\end{figure}

At the value of parameter $a=1/2$ the mono-scaling property is restored, and the spatial densities of energy $W(\xi)$ presented in Fig.~\ref{fig:6} at two sizes of the active domain practically coincide. We note that here the profiles $W(\nu)$ and $w(\xi)$ are the same because the density $\rho(\xi)$ is constant (cf.~Fig.~\ref{fig:4}). We discuss the profile at $a=1/2$ in more detail because of its simple one-hump form. The spreading of energy at $a=1/2$ is subdiffusive, with the exponent $\approx 0.47$ (see Fig.~\ref{fig:2}). This allows a tempting attempt of a simple phenomenological model for such a spreading. Indeed, the nonlinear diffusion equation 
\begin{equation}
\partial_t u(x,t)=\partial_{xx} u^{c+1},\quad c>0\;,
\label{eq:nld}
\end{equation}
describes a subdiffusive spreading of the quantity $u$, the integral of which is conserved. The self-similar solution has the scaling form $u\sim t^{-\beta}(1-4 (x/L(t))^2)^\alpha$. This solution has finite support $|x| \leq L/2$ with $L(t)\sim t^\beta$, the power $\beta<1/2$ is the single exponent of the subdiffusive spreading. The exponents $\alpha,\beta$ are determined by the nonlinearity parameter $c$: $\beta=1/(c+2)$, $\alpha=1/c$. Remarkably, the shape of the energy density profile observed numerically at $a=1/2$ is very well fitted by the form predicted by the solution of the nonlinear diffusion equation (see the black dashed curve in Fig.~\ref{fig:6} which is the function $w_{fit}(\xi)=A(1-4\xi^2)^B$ with $A\approx 1.94$ and $B\approx 2.2$). The powers, however, do not satisfy the relation $\alpha^{-1}=\beta^{-1}-2$ predicted by the nonlinear diffusion equation. Indeed, the observed value for the spreading exponent  $\beta=0.47$ corresponds to $c\approx 0.128$, but for this value of $c$, the power of the profile shape of Eq.~\eqref{eq:nld} should be $\alpha\approx 7.8$, which is very far from the fitted value $2.2$. 

\textit{Discussion.} Here, we compare the properties of the HPC and the HPG models with other cases,
where milti-scaling anomalous spreading has been observed. A typical situation of multi-scaling in anomalous diffusion, explored in Refs.~\cite{pikovsky1991statistical,castiglione1999strong}, is that of a combination of a diffusive process with ballistic modes. The latter modes describe the spreading of the support of the distribution with constant speed, so that the moment-based exponent tends to $1$ for high-order moments: $\gamma_p\to 1$ as $p\to\infty$. Furthermore, typical for such situations is a piecewise-linear shape of the dependence $p\gamma_p$ on $p$~\cite{pikovsky1991statistical,castiglione1999strong}. This type of energy spreading has been observed in Refs.~\cite{delfini2007energy,piazza2009heat,Lepri-Shchilling-Aubri-10}. However, in these studies, one considered the spreading of an energy hump on top of an equilibrium state with finite energy density. Correspondingly, the ballistic mode can be associated with sound waves~\cite{delfini2007energy}. In contradistinction, in our case of spreading of energy on top of zero-temperature equilibrium, the leading ``modes'' defining propagation of the active domain are not ballistic but super- or sub-diffusive. Correspondingly, the shape of the dependence of $p\gamma_p$ on $p$ is a smooth curve (Fig.~\ref{fig:3}) rather than a piecewise-linear line conjectured in \cite{castiglione1999strong}.

\textit{Conclusions.} In summary, we have demonstrated multi-scaled energy spreading from a local in space disturbance in the simple model of a hard-particle chain \eqref{eq:1}. Different effective lengths of the spreading domain, defined via moments of different orders or different Renyi entropies of the distribution of energies, depend on time via power laws with different, index-dependent exponents.  Correspondingly, the shapes of the energy distributions at different times are not self-similar but show a clear separation in the central peak and tails, obeying different scalings. Contrary to previous cases where ballistically spreading tails have been reported, here, the tails spread superdiffusively. 

In two remarkable limits, mono-scaling is observed, with a self-similar behavior of the energy profile. For a hard-particle gas, which corresponds to a single-well coupling potential, we observe superdiffusive spreading with exponent $\approx 2/3$ following simple scaling arguments. Here, however, the profile is nontrivial: the particle density has a minimum at the center of the domain and maxima at the edges, thus building superdiffusively spreading shocks.  The energy per particle density has a weak maximum in the middle of the domain, and at the borders, the energy density also reaches maxima. 

Another special case is a symmetric hard-particle chain with the mean distance between the particles being exactly half the potential's width. Here, in equilibrium, the pressure of the gas vanishes. We observe here a constant density of the subdiffusively (exponent $\approx 0.47$) spreading domain. The distribution of the energies is a one-hump profile very much resembling a profile of a self-similar solution of a nonlinear diffusion equation. However, the profile shape and the spreading exponent do not follow the relation resulting from the diffusion equation solution. 

An intriguing question is whether the multi-scaling observed can also occur in smooth or continuous potentials. For example, in the hard-particle gas limit $a=0$, one can replace hard particles with a more realistic model of colliding elastic spheres, interacting via Hertz law~\cite{Martinez-16,kim2018direct,PhysRevE.99.032211}.

\acknowledgments
\textit{Acknowledgments.} I thank Ph. Rosenau for useful discussions.
%

\end{document}